\documentclass{article}

\usepackage{arxiv}

\usepackage[utf8]{inputenc} 

\usepackage{amsfonts}       
\usepackage{nicefrac}       
\usepackage{microtype}      
\usepackage{lipsum}

\usepackage{color}
\usepackage{listings}
\usepackage{booktabs}
\usepackage{array}
\usepackage{graphicx}
\usepackage{longtable}
\usepackage{subfigure}

\usepackage{cite}
\usepackage{url}
\usepackage{fancyhdr}

\usepackage{soul}

\usepackage{float}
\usepackage{listings}
\usepackage{mdwmath}
\usepackage{mdwtab}
\usepackage{multirow}
\usepackage{multicol}
\usepackage{rotating}
\usepackage{setspace}
\usepackage[utf8]{inputenc}
\usepackage{lineno}
\usepackage{listings}

\usepackage{mdwmath}
\usepackage{mdwtab}
\usepackage{multirow}
\usepackage{multicol}
\usepackage{array}
\usepackage{booktabs}

\usepackage{enumitem}
\usepackage{xspace}
\usepackage[export]{adjustbox}
\usepackage{graphicx}
\usepackage{color,soul}
\usepackage{rotating}
\usepackage{setspace}
\usepackage{amsmath} 
\usepackage{amssymb}
\usepackage{float}
\usepackage{xcolor}

\usepackage{hyperref}
\usepackage[numbers]{natbib}
\usepackage{url}

\title{BGRA: A Reference Architecture for Blockchain Governance}

\author{Yue Liu\textsuperscript{1,2}, Qinghua Lu\textsuperscript{1,2}, Guangsheng Yu\textsuperscript{1}, Hye-Young Paik\textsuperscript{2}, Liming Zhu\textsuperscript{1,2}\\
\textsuperscript{1}Data61, CSIRO, Australia\\
\textsuperscript{2}University of New South Wales, Australia\\
yue.liu@data61.csiro.au, qinghua.lu@data61.csiro.au,
saber.yu@data61.csiro.au\\
h.paik@unsw.edu.au,
liming.zhu@data61.csiro.au}

\begin{document}

\maketitle

\begin{abstract}
Blockchain technology has been integrated into diverse software applications by enabling a decentralised architecture design. However, the defects of on-chain algorithmic mechanisms, and tedious disputes and debates in off-chain communities may affect the operation of blockchain systems. Accordingly, blockchain governance has received great interest for supporting the design, use, and maintenance of blockchain systems, hence improving the overall trustworthiness. Although much effort has been put into this research topic, there is a distinct lack of consideration for blockchain governance from the perspective of software architecture design. In this study, we propose a pattern-oriented reference architecture for governance-driven blockchain systems, which can provide guidance for future blockchain architecture design. We design the reference architecture based on an extensive review of architecture patterns for blockchain governance in academic literature and industry implementation. The reference architecture consists of four layers. We demonstrate the components in each layer, annotating with the identified patterns. A qualitative analysis of mapping two concrete blockchain architectures, Polkadot and Quorum, on the reference architecture is conducted, to evaluate the correctness and utility of proposed reference architecture.

\end{abstract}

Software engineering, reference architecture, blockchain governance, decision rights, incentive, accountability, pattern.

\section{Introduction}

Blockchain can provide both distributed data storage and computing platform, where a large network of untrusted participants need to reach agreements on transactional data states~\cite{scheuermann2015iacr}. As an innovative distributed ledger technology, blockchain has improved certain software attributes and brought its distinctive features, e.g., transparency, immutability, and on-chain autonomy, into various application scenarios. In recent years, blockchain has been leveraged as a software component in application systems in a substantial number of projects by enabling a decentralised infrastructure~\cite{2019-Bratanova-ACS}, for instance, energy supply~\cite{energy}, industrial IoT~\cite{IIoT}, etc.

Despite being considered a viable solution to re-architect application systems, there are significantly increased concerns that blockchain systems may suffer from the defects of on-chain algorithmic mechanisms, and tedious disputes and debates in off-chain communities. The negative crises in two world-renowned blockchain systems, Ethereum and Bitcoin, have severely affected the trustworthiness of blockchain. In 2016, the ``DAO" (Decentralised Autonomous Organisation) attack in Ethereum was caused by flaws in smart contract code and resulted in the loss of over 60 million US dollars. This was remedied by conducting a hard fork to reverse the impacted transactions~\cite{DAOattack}. While in Bitcoin, the debate of whether to increase block size caused the split of the whole ecosystem~\cite{BitcoinSize}. After these events, the blockchain community started to explore a more trustworthy governance process for both on-chain and off-chain businesses.

Blockchain governance refers to the structures and processes that are designed to ensure the development and use of blockchain are compliant with legal regulations and ethical responsibilities~\cite{liu2021systematic}. It determines the allocation of decision rights, incentives, and accountability based on the blockchain decentralisation level, which further regulates stakeholders' behaviour throughout the whole blockchain development lifecycle, and the overall blockchain ecosystem. Blockchain governance can refer to existing governance frameworks (e.g., IT governance~\cite{weill2004governance, cobit2012business}, data governance~\cite{ballard2014ibm, ISO38505}), while further investigation is also necessary considering the absence of a clear source of authority in blockchain systems. In recent years, there are continuously increasing attention focusing on this research topic, including the customised governance methods in permissioned blockchains~\cite{selected1, selected14}, regulations for blockchain-based decentralised finance (e.g., cryptocurrencies)~\cite{selected3, selected16}, etc.

Nevertheless, it is found that there is a lack of consideration for software architecture design in this area, which may hinder the design and implementation of blockchain with proper governance solutions, resulting in conflicts between stakeholders and failures of blockchain systems. In this regard, this paper presents a pattern-oriented reference architecture, which can serve as a guideline to assist system architects and developers in the development of governance-driven blockchain systems, with reusable patterns as architecture components. 

The contributions of this paper are as follows:

\begin{itemize}
    \item We propose a reference architecture to guide and facilitate the design and development of governance-driven blockchain systems. To the best of our knowledge, this is the first study learning blockchain governance from the perspective of architecture design.
    
    \item We associate multiple architectural patterns with the different components in the proposed reference architecture, to address the recurring governance-related issues in blockchain systems. The architectural patterns are gathered and analysed via a systematic literature review and further review of multiple blockchain systems.
    
    \item We evaluate the correctness and utility of our proposed reference architecture, by mapping two blockchain system architectures on the proposed reference architecture.
    
\end{itemize}

The remainder of this paper is organised as follows. Section~\ref{sec:background} introduces background knowledge and related work. Section~\ref{sec:methodology} explains our research methodology. Section~\ref{sec:architecture} presents the overall reference architecture with annotated patterns. Section~\ref{sec:evaluation} evaluates our architecture. Section \ref{sec:conclusion} concludes the paper and outlines future work.

\section{Background and Related Work}
\label{sec:background}

\subsection{Blockchain}

Blockchain was popularised by Bitcoin~\cite{Satoshi:bitcoin} and the subsequent cryptocurrencies. The concept of blockchain was then generalised to distributed ledger technology, and considered an emerging paradigm for building next-generation applications in a decentralised way. In a software system, the blockchain component can provide two core elements: (i) a distributed ledger, and (ii) a decentralised ``compute" infrastructure.

Blockchain can verify and store digital transactions via the underlying distributed ledger, without relying on any central authority to establish trust between the interoperating entities~\cite{scheuermann2015iacr}. In permissionless blockchains, trust is preserved via game theoretic incentives to maintain a majority of honest nodes~\cite{Satoshi:bitcoin}. While in permissioned blockchains, trust is achieved through the compulsory identity verification of participating entities. On-chain transactions carry the changing states of data, and blocks are containers for storing transactions. Except for the genesis block, all the blocks are linked to the previous block and thus form a chain.

Blockchain can be leveraged as a ``compute" infrastructure via the on-chain programmability (i.e., smart contracts). Smart contracts are user-defined programs that can be deployed and executed in a blockchain system to enable complex business logic such as triggers and conditions~\cite{Omohundro:2014}. For instance, Ethereum provides a built-in Turing-complete scripting language, Solidity, for developing smart contracts.

\subsection{Blockchain Governance}

Existing studies have analysed the governance frameworks for blockchain~\cite{selected11, selected14, hofman2021blockchain, liu2021defining}. Please note that in this study, the term ``blockchain governance" refers to ``governance of blockchain", where we focus on how governance fits in the development and use of blockchain.

Essentially, blockchain can be classified into three types to meet different requirements. Different types of blockchain reflect the different levels of decentralisation, and affect the governance structure regarding the allocation of decision rights, accountability, and incentives. In a blockchain system, a transparent decision-making process can help oversee whether decisions are reasonable, hence, to gain the trust of all stakeholders. Accountability can be established via both institutional and technical manners, to ensure the identifiability and answerability of stakeholders for their decisions. In blockchain governance, incentives are considered factors that may influence stakeholders' behaviours. The governance structure can provide incentives to motivate desirable behaviours and resolve conflicts between stakeholders. 

In addition, blockchain governance should be realised throughout the overall ecosystem. For on-chain transactions, the governance emphasises accountable access control. The capabilities of sending, validating, and reading transactions are assigned to different stakeholders considering the selected blockchain type. Regarding the blockchain platforms, they need to undergo a series of formalised procedures to finalise improvement proposals. Further, blockchain-based applications need to comply with industry regulations and specifications, any changes may lead to upgrades of the underlying blockchain platform. For the off-chain community governance, the stakeholders are gradually divided into different groups regarding their roles and decision rights, e.g., stakeholders may have different communication channels for certain issues. 

Furthermore, blockchain governance should ensure that the related decisions and processes conform to legal regulations and ethical responsibilities. Specifically, managing legal compliance relies on local and international policies regarding where to deploy a blockchain, while promoting ethical guidelines can preserve human values in blockchain governance.

In recent years, many researchers explore the topic of blockchain governance from diverse perspectives. For instance, Katina et al.~\cite{selected5} propose and analyse seven interrelated elements of blockchain governance, including philosophy, theory, axiology, methodology, axiomatic, method and applications. Beck et al.~\cite{selected14} adopt the three major dimensions of IT governance (i.e., decision rights, incentives, and accountability), and discuss their allocations in blockchain governance. Allen and Berg~\cite{selected7} focus on the exogenous and endogenous governance methods for blockchain platforms, while John and Pam~\cite{selected10} and Pelt et al.~\cite{selected11} both investigate this topic regarding on-chain and off-chain development processes. Hofman et al.~\cite{hofman2021blockchain} propose a high-level analytic framework for blockchain governance, covering six different aspects (i.e., why, who, when, what, where, and how). However, the existing studies only provide general discussion and high-level principles to realise governance, while practitioners require more detailed solutions to face the learning curve and facilitate the architecture design of governance-driven blockchain systems.


\begin{figure*}[t]
	\centering
	\includegraphics[width=0.9\columnwidth]{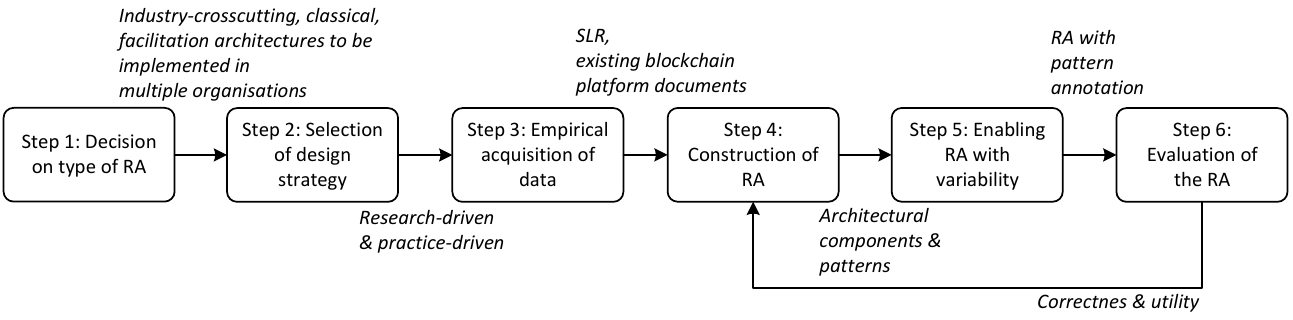}	
	\caption{Methodology.}
	\label{fig:methodology}
\end{figure*}

\subsection{Reference Architecture}

A reference architecture can be regarded as ``a reference model mapped onto software elements (that cooperatively implement the functionality defined in the reference model) and the data flow between them. Whereas a reference model divides the functionality, a reference architecture is the mapping of that functionality onto a system decomposition"~\cite{bass2003software}.  A reference architecture can support system development by addressing the inclusive business rules, architectural styles, best practices of software development, and software elements~\cite{RA_concept}. There are existing studies about blockchain reference model and reference architecture. 
For instance, Yuan and Wang propose a 6-layer reference model of blockchain~\cite{FeiYue_RA}, while Ellervee et al.~\cite{ellervee2017comprehensive} present a blockchain reference model from the perspectives of actors, services, processes, and data models. Homoliak et al.~\cite{security_RA} present a security reference architecture for blockchain based on the blockchain network implementation stacks proposed by Wang et al.~\cite{BC_stack_model}. In addition, there are reference architectures for diverse blockchain-based applications, for instance, crowdsourcing~\cite{Crowdsourcing_RA}, healthcare~\cite{Healthcare_RA}, and government services~\cite{Government_RA}, etc. Nonetheless, we found a lack of consideration of software architecture design for blockchain governance. Hence, this study illustrates a pattern-oriented reference architecture for governance-driven blockchain systems.

\section{Methodology}
\label{sec:methodology}

This section introduces the methodology of this study. We adopted an empirically-grounded design methodology for reference architecture~\cite{RA_methodology}, and the overall research process is illustrated in Figure~\ref{fig:methodology}. There are six steps, and each step has an output to the following step.

The first step is to determine the type of our reference architecture. Based on Galster and Avgeriou's proposal~\cite{RA_methodology}, the reference architecture is an industry-crosscutting (\textit{usage context}), classical (\textit{when}), facilitation (\textit{why}) architecture to be implemented in multiple organisations (\textit{where}). Hereby, ``industry-crosscutting" means that the reference architecture can cover more than one industry, and “classical" refers to that this reference architecture is developed based on existing blockchain systems. ``Facilitation" indicates that the reference architecture aims to provide guidance for the future design of blockchain systems, while “multiple organisations" is determined by the decentralised nature of blockchain.


The second step is to select our design strategy. In this study, our design strategy is a combination of ``research-driven" and ``practice-driven". ``Research-driven" means that the design of this reference architecture is based on state-of-the-art research from a systematic literature review, while ``practice-driven" is also applicable since this study is based on the review results of multiple extant blockchain systems and summary of the best-practices for blockchain governance.



The third step is the empirical acquisition of data. This study adopts and extends several existing studies as data acquisition. Specifically, Liu et al. performed a systematic literature review, in which 37 primary studies were selected and analysed regarding six research questions~\cite{liu2021systematic}. The extracted results covered the definition, motivations, objects, process, stakeholders, and mechanisms of blockchain governance. Afterwards, the researchers reviewed the existing governance frameworks and standards (i.e., IT governance~\cite{weill2004governance, cobit2012business}, data governance~\cite{ballard2014ibm, ISO38505}, OSS governance~\cite{o2007emergence, de2007governance}, platform ecosystem governance~\cite{tiwana2010platform}), to understand the characteristics of blockchain governance. They also scrutinised the open websites and documents of five blockchain platforms (i.e., Bitcoin, Ethereum, Dash, Tezos, and Hyperledger Fabric), to understand how blockchain governance is implemented in a real-world context~\cite{liu2021defining}. In addition, they presented a pattern language for blockchain governance~\cite{pattern_collection}.


Based on the acquired data, in the next step, we constructed a reference architecture for governance-driven blockchain systems by integrating the architectural patterns into a widely-accepted reference model of blockchain~\cite{ISO23257}. Meanwhile, the variability of our reference architecture design in step five was enabled by annotating that applying different patterns can lead to the instantiation of various concrete architectures for blockchain systems.

In the final step, the evaluation of our proposed reference architecture was carried out by reviewing two other blockchain systems, Polkadot and Quorum. We map the architectural components of Polkadot and Quorum on the proposed reference architecture, to prove that the reference architecture can be transformed into meaningful concrete architectures. Further, the evaluation results can help refine the reference architecture.



\begin{figure*}[t]
	\centering
	\includegraphics[width=0.7\textwidth]{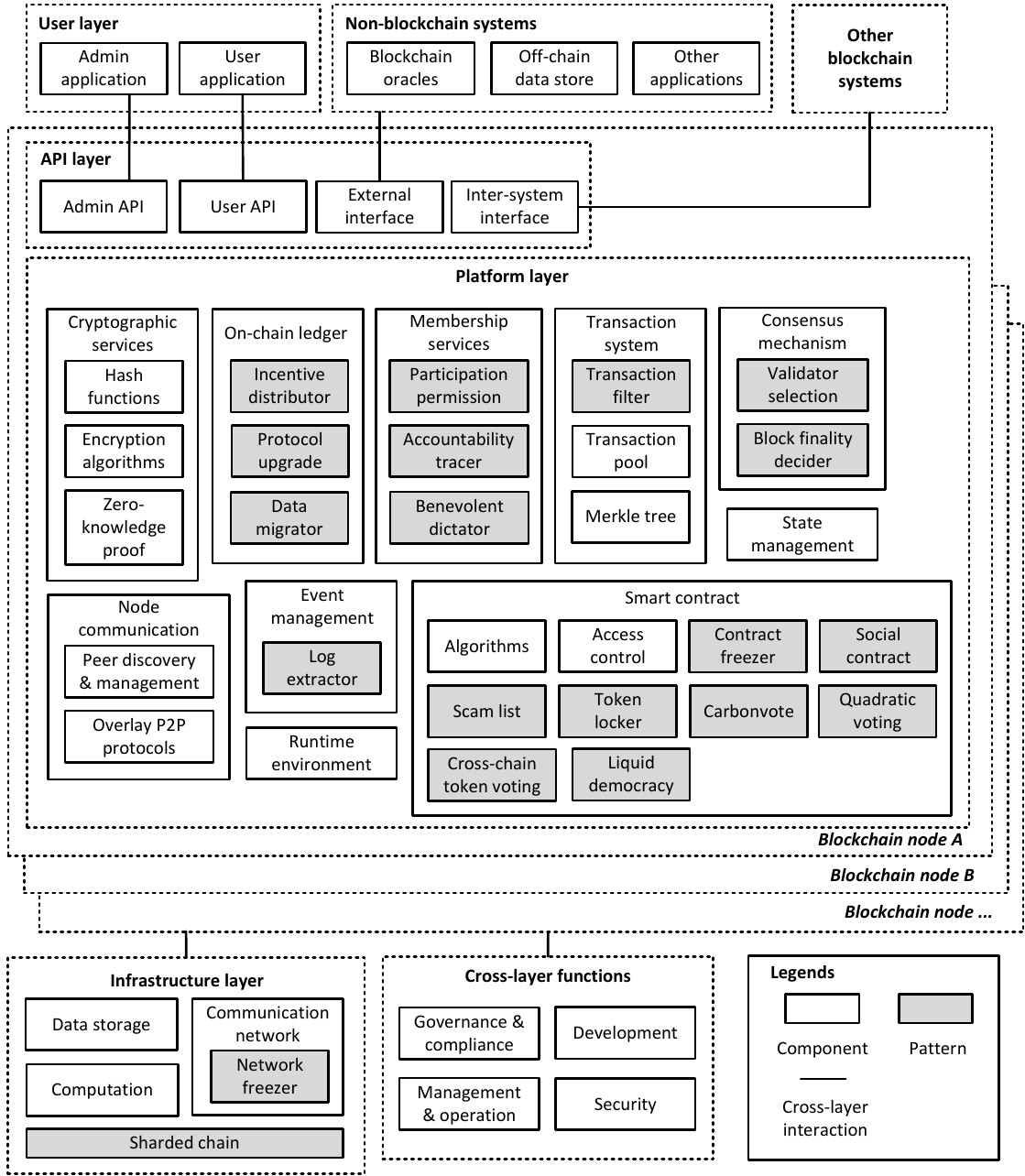}	
	\caption{A pattern-oriented reference architecture for governance-driven blockchain systems.}
	\label{fig:architecture}
\end{figure*}

\section{Reference Architecture}
\label{sec:architecture}

In this section, we present a pattern-oriented reference architecture for governance-driven blockchain systems. Figure~\ref{fig:architecture} illustrates the overview of the architecture, which consists of: 1) infrastructure layer, 2) platform layer, 3) API layer, 4) user layer, and 5) cross-layer functions. Specifically, the platform layer and API layer should be implemented in each participating node, and all nodes in a blockchain system share the same infrastructure layer and cross-layer functions. Moreover, this figure includes non-blockchain systems and other blockchain systems to illustrate the interactions of API layer. We apply a set of patterns as architectural components to realise governance in the reference architecture, which are annotated in the figure. In addition, we summarise these components in Table~\ref{tab:components}, to explain the applicable decentralised level, type, and responsibility of each annotated component.

\begin{table*}[tbp]
\footnotesize
\centering
\caption{Pattern-oriented components in the reference architecture.}
\label{tab:components}
\begin{tabular}{p{0.14\columnwidth}p{0.15\columnwidth}p{0.075\columnwidth}p{0.585\columnwidth}}
\toprule

{\bf Component} &
{\bf  Decentralisation level} &
{\bf  Type} &
\multicolumn{1}{c}{\bf Responsibility}\\
\midrule

\multirow{2}{0.14\columnwidth}{Network freezer} & \multirow{2}{0.15\columnwidth}{Permissioned \& permissionless} & \multirow{2}{0.2\columnwidth}{Optional} & \multirow{2}{0.585\columnwidth}{Suspending blockchain transactions at the network level, to stop the broadcast of malicious transactions.}
\\
\cmidrule(l){1-4}

\multirow{3}{0.14\columnwidth}{Sharded chain} & \multirow{3}{0.15\columnwidth}{Permissioned \& permissionless} & \multirow{3}{0.2\columnwidth}{Optional} & \multirow{3}{0.585\columnwidth}{Splitting blockchain into multiple shards, where the data storage, computation, and communication are accordingly split, to improve the scalability of blockchain systems.}
\\ \\
\cmidrule(l){1-4}

\multirow{2}{0.14\columnwidth}{Incentive distributor} & \multirow{2}{0.15\columnwidth}{Permissioned \& permissionless} &  \multirow{2}{0.2\columnwidth}{Optional} & \multirow{2}{0.585\columnwidth}{Providing on-chain tokens to drive the motivation and behaviour of stakeholders in decision-making process.}
\\
\cmidrule(l){1-4}

\multirow{2}{0.14\columnwidth}{Protocol upgrade} & \multirow{2}{0.15\columnwidth}{Permissioned \& permissionless} & \multirow{2}{0.2\columnwidth}{Mandatory} & \multirow{2}{0.585\columnwidth}{Implementing the software upgrades to a blockchain system.}
\\
\cmidrule(l){1-4}

\multirow{2}{0.14\columnwidth}{Data migrator} & \multirow{2}{0.15\columnwidth}{Permissioned \& permissionless} & \multirow{2}{0.2\columnwidth}{Optional} & \multirow{2}{0.585\columnwidth}{Migrating data from a source blockchain system to target blockchain system(s) based on data governance and management requirements.}
\\
\cmidrule(l){1-4}

\multirow{2}{0.14\columnwidth}{Participation permission} & \multirow{2}{0.15\columnwidth}{Permissioned} & \multirow{2}{0.2\columnwidth}{Optional} & \multirow{2}{0.585\columnwidth}{Managing the participation to a blockchain system, requiring real-world identity verification and the approval of authorities.}
\\
\cmidrule(l){1-4}

\multirow{2}{0.14\columnwidth}{Accountability tracer} & \multirow{2}{0.15\columnwidth}{Permissioned \& permissionless} &  \multirow{2}{0.2\columnwidth}{Mandatory} & \multirow{2}{0.585\columnwidth}{Identifying the source of a blockchain transaction, to ensure the accountability of transaction senders.}
\\
\cmidrule(l){1-4}

\multirow{2}{0.14\columnwidth}{Benevolent dictator} & \multirow{2}{0.15\columnwidth}{Permissioned \& permissionless} & \multirow{2}{0.2\columnwidth}{Mandatory} & \multirow{2}{0.585\columnwidth}{Specific stakeholders possess additional decision rights for certain governance-related issues.}
\\
\cmidrule(l){1-4}

\multirow{2}{0.14\columnwidth}{Transaction filter} & \multirow{2}{0.15\columnwidth}{Permissioned \& permissionless} & \multirow{2}{0.2\columnwidth}{Optional} & \multirow{2}{0.585\columnwidth}{Examining submitted transactions to ensure the validity of transaction format/content, rejecting and discarding invalid transactions.}
\\
\cmidrule(l){1-4}

\multirow{2}{0.14\columnwidth}{Validator selection} & \multirow{2}{0.15\columnwidth}{Permissioned \& permissionless} &  \multirow{2}{0.2\columnwidth}{Mandatory} & \multirow{2}{0.585\columnwidth}{Selecting the node operator who is eligible to validate and append the candidate block to the blockchain.}
\\
\cmidrule(l){1-4}

\multirow{2}{0.14\columnwidth}{Block finality decider} & \multirow{2}{0.15\columnwidth}{Permissioned \& permissionless} &  \multirow{2}{0.2\columnwidth}{Optional} & \multirow{2}{0.585\columnwidth}{Waiting for a certain number of subsequent blocks to confirm that a previous block and its contained data is finalised and immutable.}
\\
\cmidrule(l){1-4}

\multirow{2}{0.14\columnwidth}{Log extractor} & \multirow{2}{0.15\columnwidth}{Permissioned \& permissionless} & \multirow{2}{0.2\columnwidth}{Optional} & \multirow{2}{0.585\columnwidth}{Extracting logged event information from the blockchain system for further analysis and audit.}
\\
\cmidrule(l){1-4}

\multirow{2}{0.14\columnwidth}{Contract freezer} & \multirow{2}{0.15\columnwidth}{Permissioned \& permissionless} & \multirow{2}{0.2\columnwidth}{Optional} & \multirow{2}{0.585\columnwidth}{Suspending all the operations to a particular smart contract.}
\\
\cmidrule(l){1-4}

\multirow{2}{0.14\columnwidth}{Social contract} & \multirow{2}{0.15\columnwidth}{Permissioned \& permissionless} & \multirow{2}{0.2\columnwidth}{Optional} & \multirow{2}{0.585\columnwidth}{Specifying the future maintainer or qualification of maintainers for a blockchain system.}
\\
\cmidrule(l){1-4}

Scam list & Permissionless & Optional & Listing the malicious blockchain addresses to warn all stakeholders of risky interactions.
\\
\cmidrule(l){1-4}

\multirow{2}{0.14\columnwidth}{Token locker} & \multirow{2}{0.15\columnwidth}{Permissionless} & \multirow{2}{0.2\columnwidth}{Optional} & 
Locking a certain amount of on-chain tokens for a specified time period, to restrict the token holder's behaviour in a decision-making process.
\\
\cmidrule(l){1-4}


\multirow{2}{0.14\columnwidth}{Carbonvote} & \multirow{2}{0.15\columnwidth}{Permissionless} & \multirow{2}{0.2\columnwidth}{Optional} & Counting votes for improvement proposals according to the tokens held by blockchain addresses, to prevent Sybil attack.
\\
\cmidrule(l){1-4}

\multirow{2}{0.14\columnwidth}{Quadratic voting} & \multirow{2}{0.15\columnwidth}{Permissionless} & \multirow{2}{0.2\columnwidth}{Optional} & Consuming $n^{2}$ number of tokens when a blockchain address submits $n$ votes for an improvement proposal, to capture the preference of stakeholders' decisions.
\\
\cmidrule(l){1-4}

Cross-chain token voting & \multirow{2}{0.15\columnwidth}{Permissionless} & \multirow{2}{0.2\columnwidth}{Optional} & Issuing tokens and holding votes in other blockchain systems, for the improvement proposals in the original blockchain system.
\\
\cmidrule(l){1-4}

\multirow{2}{0.14\columnwidth}{Liquid democracy} & \multirow{2}{0.15\columnwidth}{Permissionless} & \multirow{2}{0.2\columnwidth}{Optional} & Delegating the decision rights and revoking the delegation for improvement proposals to/from other stakeholders.
\\

\bottomrule
\end{tabular}
\end{table*}

\subsection{Infrastructure Layer}
First, the infrastructure layer of a blockchain system consists of the functional components for building a fundamental operating environment, including \textit{data storage}, \textit{communication network}, and \textit{computation}. For \textit{data storage}, the blockchain system architects or developers need to determine the physical and logical location of on-chain data, and corresponding CRUD (i.e., create, read, update and delete) operations. \textit{Communication network} includes the peer-to-peer network for blockchain nodes, connection between stakeholders and the blockchain system, and interactions between a blockchain system with other systems. In the \textit{communication network}, \textit{\textbf{network freezer}} can be leveraged by the system administrators or governors to suspend all on-chain business. This pattern can disconnect the nodes or block data traffic in a short time, to avoid the negative impact in an emergency situation. A frozen blockchain system requires human interventions for reactivation. \textit{Computation} enables the runtime environment for each node, and on-chain programmability with complex business logic. In this layer, the structure of blockchain may influence the other three main components. Specifically, a blockchain can have either a single or multiple shards. A \textit{\textbf{sharded chain}} refers to that a blockchain is partitioned into different shards, consequently, the data storage, communication network, and computation are accordingly split regarding the shards. Blockchain nodes only need to process the transactions in their own shards. Compared with single-shard blockchains, a \textit{\textbf{sharded chain}} is more scalable and has better performance.

\subsection{Platform Layer}
The core services and features of a blockchain system are implemented and embodied in the platform layer. The main components in this layer include \textit{cryptographic services}, \textit{on-chain ledger}, \textit{membership services}, \textit{transaction system}, \textit{consensus mechanism}, \textit{node communication}, \textit{event management}, \textit{runtime environment}, and \textit{smart contract}.

A blockchain system incorporates a series of \textit{cryptographic services} to preserve data confidentiality and integrity. For instance, \textit{hash functions} can map arbitrary-size data to fixed-size data. The above-mentioned \textit{Merkle tree} is supported by \textit{hash functions}. \textit{Encryption algorithms} can generate and decrypt ciphertext via secret keys. \textit{Zero-knowledge proof} can preserve privacy in verification issues by only confirming that an entity knows or possess certain data without revealing the actual data. 

Essentially, the \textit{on-chain ledger} is the implementation of \textit{data storage} in the infrastructure layer. Each participating node maintains a local replica of the blockchain ledger to preserve data integrity, availability, and consistency. A full node keeps all historical transaction information while a light node can only store the block headers, but a light node needs to rely on full nodes for data enquiry. To align the distinct objectives of stakeholders, especially the nodes, the \textit{\textbf{incentive distributor}} rewards tokens (i.e., programmable digital assets) to stakeholders who obey the codified rules and contribute to the operation of a blockchain system. More generally, the \textit{\textbf{incentive distributor}} can drive stakeholders' motivation and behaviour in a decision-making process. In addition, new functionalities of the blockchain system are implemented via \textit{\textbf{protocol upgrade}}, which may affect the records of on-chain ledger if forking is required. Please note that there are two types of forking: i) backward-compatible upgrades as soft forks, and ii) backward-incompatible upgrades as hard forks. For on-chain ledger data, a \textit{\textbf{data migrator}} is responsible for interacting with external migration tools, which can help migrate the ledger data from a source blockchain system to a target blockchain system, enabling more comprehensive on-chain data management and governance services. Migrating on-chain data can be realised via multiple ways~\cite{data_migration}, for instance, generating a snapshot of the source blockchain (including the entire states, smart contracts, and transactions), cloning a node from the source blockchain system, etc.

\textit{Membership services} implemented in blockchain systems are related to the decentralisation level of the deployed blockchain. Specifically, \textit{\textbf{participation permission}} refers to the identity verification of stakeholders, and approval of authorities (e.g., system administrator) in permissioned blockchain systems. \textit{\textbf{Accountability tracer}} is enabled by the digital signature of transaction senders. Every transaction needs the sender's signature, which is generated via two steps: i) hashing the original data, and ii) encrypting the hash value. Transactions need to be verified by block validators before they are officially recorded. If a transaction contains malicious information, the decrypted hash value can be used to check whether the transaction data is altered during transmission, and the signature can ensure the traceability and identifiability of transaction senders. Please note that in permissioned blockchain systems, accountability can be realised in terms of identifying the real-world stakeholders, while in permissionless blockchain systems, only accountable blockchain addresses can be located due to the inherent anonymity. In permissionless blockchain systems, \textit{\textbf{benevolent dictator}} is arranged throughout the blockchain development and operation stages, referring to stakeholders who have more decision rights than others. For instance, stakeholders may trust the decisions of core developers, who are considered the benevolent dictators based on their expertise of technical meritocracy and the collective benefits of update decisions.

The \textit{transaction system} is closely connected to the \textit{on-chain ledger}. This component can be regarded as the data entry of blockchain. All varying data states on blockchain are carried by the transactions. When a transaction is generated and sent to the blockchain system, a deployed \textit{\textbf{transaction filter}} can examine whether the submitted transactions meet the format or content requirements predefined by the blockchain project team or administrators, to avoid unauthorised or harmful information being fed to blockchain. The valid transactions are temporarily collected in the \textit{transaction pool}, which is a local memory in each node. Nodes can select transactions from the pool and generate candidate blocks, in which the transaction information is compressed in the form of \textit{Merkle tree}. A \textit{Merkle tree} is created via hashing the transactions, and then iteratively summarising and hashing the hash values until a Merkle root is generated. This data structure can preserve data integrity, facilitate the verification of historical transactions, and save the local space of nodes.

Blockchain in practice is a distributed ledger technology where each participating node holds a local replica of the whole ledger contents. A critical issue is how the multiple nodes can agree with the states of blockchain. Conflicts about the blockchain states will impact the security and availability of the overall platform: i) The ledger contents may be compromised if they are not synchronised across the nodes. Attackers can easily modify historical transactions and claim to be the valid version. ii) Requests for the same data at the same time are replied with identical responses. To address the above problems, \textit{consensus mechanisms} are leveraged as a governance method to align the agreement of different nodes. When node operators join a blockchain system, they should all synchronise the ledger contents. During the blockchain operation, each node collects pending transactions and generates a candidate block. \textit{\textbf{Validator selection}} decides a block validator each round, who is allowed to append its block to the blockchain, while other nodes all need to synchronise this block in their local replicas. The block validator can be selected according to different criteria, which are implemented as diverse consensus mechanisms (e.g., computation capability in Proof-of-Work, possessed stakes in Proof-of-Stake, appointed by the system administrator in Proof-of-Authority, etc.). \textit{\textbf{Block finality decider}} reinforces the immutability of blocks and their contained transactions in the way that, after a block is appended to blockchain, certain numbers of subsequent blocks can be regarded as the confirmation to ensure that the previous block is recorded and finalised with high probability. 

In a blockchain system, each node needs to keep listening to the network to collect broadcast transactions and synchronise appended blocks, which is accomplished via \textit{peer discovery and management} and \textit{overlay P2P protocols}. These two components compose the \textit{node communication} component. The \textit{state management} component can be exploited to update the on-chain digital assets (e.g., tokens) based on the new transactions, while the \textit{event management} component logs the required information when a transaction triggers particular event(s). Stakeholders can analyse the logged information via \textit{\textbf{log extractor}}. In addition, \textit{runtime environment} supports the execution of smart contracts.


\textit{Smart contract} denotes the programs run in a blockchain, which enables the decentralised applications built on a blockchain system. A smart contract function is triggered by transactions, and the outputs can be stored on-chain. Smart contracts can be codified with various algorithms to provide different functionalities, or access control mechanisms to restrict users' behaviour~\cite{xu2018pattern}. Developers can implement \textit{\textbf{contract freezer}} in smart contracts, and define stakeholders who are eligible to trigger the freezer. \textit{\textbf{Contract freezer}} can pause or terminate all operations to a smart contract, when an attack is made to this contract. In addition, the blockchain project team or system administrators can deploy a special kind of smart contract, \textit{\textbf{social contract}}, to announce the future maintainers, or qualification of future maintainers for a blockchain system. If malicious operations are identified, the related blockchain addresses, both stakeholders and smart contracts, can be recorded and listed in the \textit{\textbf{scam list}}, which is also in the form of smart contract. Other stakeholders can stop interacting with these scams referring to the list. \textit{\textbf{Token locker}} can be deployed in permissionless blockchain systems, which can grant stakeholders the decision rights for specific governance issues (e.g., the approval of improvement proposals) when stakeholders lock a particular number of tokens for a certain time period as the security deposit. If stakeholders do not obey the rules during a decision-making process, the decision rights will be revoked, and the locked tokens may be destructed. In addition, a series of patterns for voting can be implemented via smart contracts to resolve conflicts and reach consensus. For instance, to update a blockchain system, stakeholders can submit, broadcast, and discuss improvement proposals via off-chain means, while the final decisions are usually made by virtue of voting. In \textit{\textbf{carbonvote}}, the votes are counted in terms of the number of tokens possessed by stakeholders, to prevent the Sybil attack where a stakeholder can register multiple blockchain addresses to compromise the vote. \textit{\textbf{Quadratic voting}} can express stakeholders' preferences when finalising improvement proposals. In this voting scheme, voting for improvement proposals consumes tokens as funds, and the preference is indicated via the exponential increase of consumed tokens that submitting $n$ votes will cost $n^2$ number of tokens. \textit{\textbf{Cross-chain token voting}} requires the interactions between blockchain systems. The blockchain project team or system administrator needs to issue tokens and deploy smart contracts for voting in source blockchain systems, and the token holders are eligible to vote for improvement proposals in the original blockchain system. \textit{\textbf{Liquid democracy}} allows stakeholders to delegate the decision rights to other trusted stakeholders, and revoke the delegation anytime. Finally, all approved improvement proposals are implemented via \textit{\textbf{protocol upgrade}}.

\subsection{API Layer, User Layer, and Other Systems}  

The API layer consists of four main types of functions for a blockchain node to provide services to different applications and systems. Specifically, \textit{admin API} and \textit{user API} can provide access to the platform layer components for system administrators and users via \textit{admin application} and \textit{user application} in the user layer respectively. The \textit{external interface} can connect a blockchain system to non-blockchain systems such as oracles, and off-chain databases, while the \textit{inter-system interface} is for the communication between nodes in different blockchain systems. 

\subsection{Cross-Layer Functions}

Cross-layer functions include \textit{governance and compliance}, \textit{development}, \textit{management and operation}, and \textit{security}, to provide auxiliary services to components in all other layers. In particular, \textit{governance and compliance} highlights the allocation of decision rights, incentives, and accountability within a blockchain system. This component specifies the high-level guidelines of how a blockchain system is operated and maintained, to meet the legal regulations, industry specifications, and broader ethical requirements, while other cross-layer functions need to refer to its guidance. The \textit{development} component is exclusively for developers' activities, for instance, codifying and testing the updated rules, building software packages, etc. This component needs to record developers' contributions for further incentive distribution and accountability process. \textit{Management and operation} can be regarded as the execution of \textit{governance and compliance} by system administrators, who need to monitor, manage, and control the blockchain system to ensure normal operation and risk management. The \textit{security} component can preserve data confidentiality, integrity and availability in a blockchain system, via predefined decision rights and fine-grained access control of certain stakeholders (e.g., node operators) to restrict their behaviour, positive or negative incentives to drive their motivations, and identity verification and management to establish a complete accountability process.

\section{Evaluation}
\label{sec:evaluation}

\begin{figure*}[t]
	\centering
	\includegraphics[width=\columnwidth]{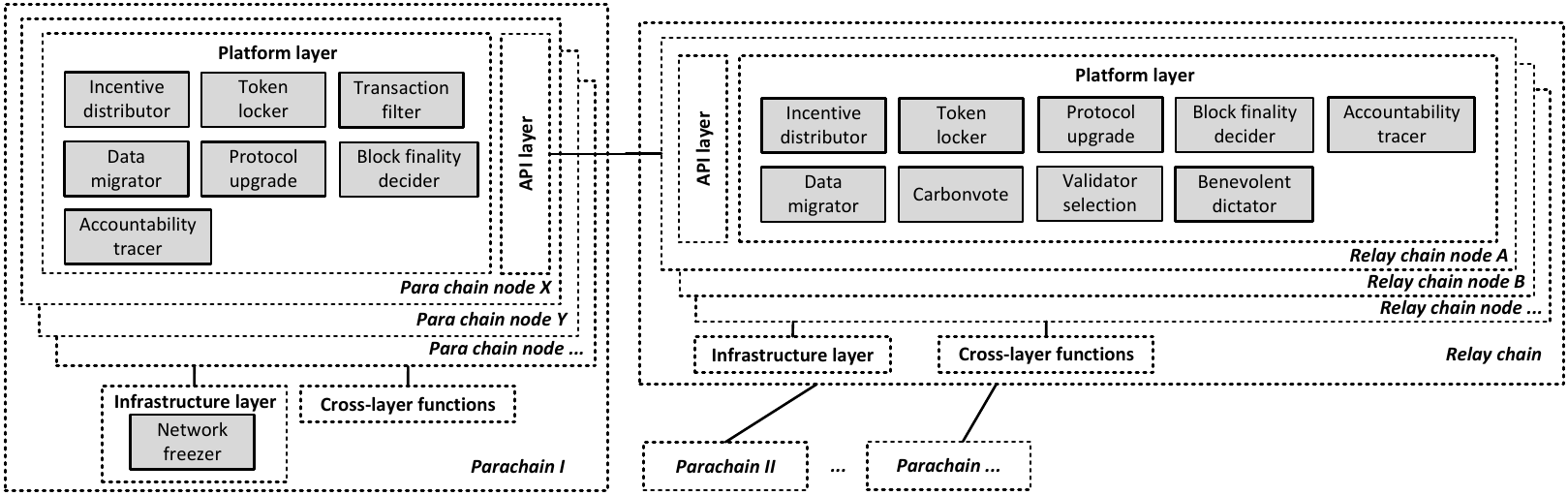}	
	\caption{Architecture mapping of Polkadot.}
	\label{fig:polkadot}
\end{figure*}

\begin{figure}[t]
	\centering
	\includegraphics[width=0.39\columnwidth]{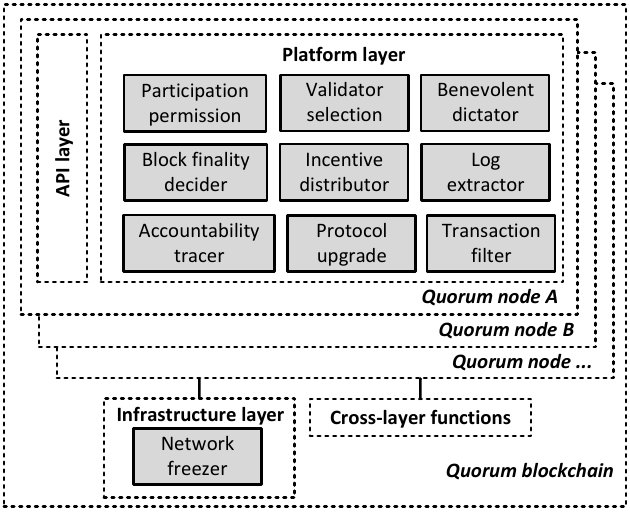}	
	\caption{Architecture mapping of Quorum.}
	\label{fig:quorum}
\end{figure}

In this section, we evaluate the correctness and utility of our proposed reference architecture by mapping existing blockchain system architectures on the reference architecture. Regarding that our reference architecture is adapted from a widely-accepted reference model, the evaluation focuses on the validation of pattern-oriented architecture components for governance. We selected two blockchain systems: Polkadot\footnote{https://polkadot.network/} and Quorum\footnote{https://consensys.net/quorum/}. Polkadot maintains a permissionless multi-chain ecosystem well-known for the cross-chain interoperability, while Quorum provides permissioned blockchain systems for enterprises and individuals. We collected and scrutinised the available documents provided by these two blockchain systems, and mapped the pattern-oriented components to the reference architecture. Figure~\ref{fig:polkadot} and \ref{fig:quorum} demonstrate the simplified architecture mapping of Polkadot and Quorum respectively, and Table~\ref{tab:comparison} present an intuitive comparison of these two blockchain systems regarding the use of components for governance.

\begin{table*}[tbp]
\footnotesize
\centering
\caption{Comparison of the use of pattern-oriented components in Polkadot and Quorum architectures.}
\label{tab:comparison}
\begin{tabular}{p{0.07\columnwidth}p{0.4\columnwidth}p{0.4\columnwidth}}
\toprule

{\bf Component} &
\multicolumn{1}{c}{\bf  Polkadot} &
\multicolumn{1}{c}{\bf  Quorum} \\
\midrule

\multirow{2}{0.07\columnwidth}{Network freezer} & Block validators can vote to suspend the validation system of a certain parachain. & System administrators can suspend the operations of blockchain nodes or accounts to freeze the system. \\
\cmidrule(l){1-3}

\multirow{2}{0.07\columnwidth}{Sharded chain} & Parachains are operated as shards, while the relay chain is regarded as the coordinator between different parachains. & \multirow{2}{0.4\columnwidth}{N/A} \\
\cmidrule(l){1-3}

\multirow{2}{0.07\columnwidth}{Incentive distributor} & Stakeholders are rewarded for their contributions to Polkadot operation. & Incentives are inactivated by default, and can be triggered during deployment. \\
\cmidrule(l){1-3}

\multirow{2}{0.07\columnwidth}{Protocol upgrade} & \multirow{2}{0.4\columnwidth}{Polkadot can upgrade the on-chain protocol without forking.} & All participants need to approve the upgrade, and update the configuration file. \\
\cmidrule(l){1-3}

\multirow{2}{0.07\columnwidth}{Data migrator} & Polkadot can interact with external blockchain systems via cross-consensus messages. & \multirow{2}{0.4\columnwidth}{N/A} \\
\cmidrule(l){1-3}

\multirow{2}{0.07\columnwidth}{Participation permission} & \multirow{2}{0.4\columnwidth}{N/A} & The deployer of a Quorum system needs to directly send invitations to other participants. \\
\cmidrule(l){1-3}

Accountabil-ity tracer & Polkadot participants are identified via their on-chain addresses. & The accountability includes stakeholders' real-world identities based on \textbf{\textit{participation permission}}. \\
\cmidrule(l){1-3}

\multirow{2}{0.07\columnwidth}{Benevolent dictator} & The council and technical committee have additional decision rights for improvement proposals. & The Quorum project team can provide additional support to a Quorum system. \\
\cmidrule(l){1-3}

\multirow{2}{0.07\columnwidth}{Transaction filter} & Polkadot defines a universal transaction format across the system. & Quorum offers different transaction types, transactions not meeting specific type requirements cannot be executed. \\
\cmidrule(l){1-3}

\multirow{2}{0.07\columnwidth}{Validator selection} & Block validators are selected according to the staked tokens of candidates and nominators. & Block validators are determined and assigned by the system administrators. \\
\cmidrule(l){1-3}

\multirow{2}{0.07\columnwidth}{Block finality decider} & Block validators vote to decide the valid chain, where the blocks are finalised. & In Quorum, certain consensus protocols can achieve immediate finality after a new block is appended to the blockchain. \\
\cmidrule(l){1-3}

\multirow{2}{0.07\columnwidth}{Log extractor} & \multirow{2}{0.4\columnwidth}{N/A} & System administrators can monitor and analyse all activities within a Quorum system. \\
\cmidrule(l){1-3}

\multirow{2}{0.07\columnwidth}{Token locker} & Acquiring certain decision rights requires the locking of Polkadot tokens, e.g., becoming a block validator. & \multirow{2}{0.4\columnwidth}{N/A} \\
\cmidrule(l){1-3}

\multirow{2}{0.07\columnwidth}{Carbonvote} & Votes are counted regarding the number and locking period of Polkadot tokens. & \multirow{2}{0.4\columnwidth}{N/A} \\

\bottomrule
\end{tabular}
\end{table*}

\subsection{Architecture mapping of Polkadot}

Polkadot consists of multiple parachains which can process transactions independently, and a relay chain to collect and confirm all blocks generated by each parachain and enable communication between them. In general, Polkadot can be regarded as a replicated sharded state machine (\textit{\textbf{sharded chain}}) where all parachains operate as shards, and the relay chain is responsible to preserve the consensus among all shards~\cite{Polkadot_design}. While inter-shard communication is facilitated by the relay chain, \textit{\textbf{data migrator}} is realised via the cross-consensus message format and protocols, through which Polkadot can send, receive and process data to/from external blockchain systems~\cite{Polkadot_xcm}.

Polkadot implements \textbf{\textit{incentive distributor}} via inherent token issuance. Stakeholders contributing to the relay chain and parachain operation are rewarded with tokens, e.g., validators and nominators can obtain tokens according to their staked tokens after block inclusion in the relay chain, and fishermen can get rewards by reporting illegal actions in parachains~\cite{Polkadot_design}. In Polkadot, transaction fees are split into two parts: one fraction is paid to the validator, while the other fraction is saved to support the implementation of future improvement proposals. Meanwhile, \textbf{\textit{token locker}} is enforced in different activities in Polkadot to assign certain decision rights to stakeholders while also restraint their behaviors~\cite{wood2016polkadot, Polkadot_design}. For instance, the selection of block validators, auction of parachain slots, and voting of improvement proposals all require stakeholders to deposit a certain number of tokens during the event. Any malicious operations may cause the loss of Polkadot tokens.

In Polkadot, all on-chain \textit{\textbf{protocol upgrades}} need to undergo referendum before implementation. A variant of \textbf{\textit{carbonvote}} is found applied in Polkadot that votes are counted regarding the number of staked tokens, and also the staked period~\cite{Polkadot_design}. A voluntary extended locking of tokens can increase the voting power of stakeholders, since a long-term locking can express the preference of stakeholders' decisions to some extent. Accepted proposals are implemented via upgrading Polkadot's WebAssembly execution host without the need of forking~\cite{Polkadot_upgrade}.

Polkadot employs the Nominated Proof-of-Stake consensus mechanism, where \textbf{\textit{validator selection}} is according to the staked tokens of validator candidates themselves or nominators~\cite{Polkadot_design}. Finally, a set of validators are selected and randomly assigned to each parachain at the beginning of every era (i.e., roughly one day). In each parachain, collators are responsible for the collection and execution of transactions, and generate blocks for the assigned validators. Note that Polkadot leverages \textbf{\textit{transaction filter}} by defining a universal transaction format~\cite{Polkadot_transaction}. Validators need to examine the parachain blocks, while a relay chain block is produced via the Blind Assignment for Blockchain Extension protocol~\cite{Polkadot_babe}. For each parachain, the \textbf{\textit{block finality decider}} would be the relay chain block, which means that a prachain block is finalised when it is included in a relay chain block. Whilst, the \textbf{\textit{block finality decider}} for Polkadot's relay chain is the GRANDPA protocol~\cite{Polkadot_grandpa}, in which validators need to vote for the longest relay chain. When more than two third of validators affirm the same chain containing a particular same block, this block and all its predecessors are finalised.

In Polkadot, the council and technical committee can be considered the \textbf{\textit{benevolent dictator}}, who have the rights to trigger fast-tracked referenda, or cancel an improvement proposal or referendum via internal voting. Within Polkadot, the \textit{\textbf{accountability tracer}} is realised via participants' on-chain addresses, and a common penalty is to destroy the staked tokens of malicious participants. Finally, validators can vote to activate the \textbf{\textit{network freezer}} of a parachain to suspend its validation system, and the recovery can be decided via either a validator-voting or referendum.

\subsection{Architecture mapping of Quorum}

Blockchain application providers can deploy Quorum blockchain systems via Blockchain as a service. Basically, Quorum blockchain systems can be considered permissioned Ethereum systems for enterprises or individuals. Consequently, \textbf{\textit{participation permission}} is realised that the deployer (i.e., blockchain application provider) can directly send invitations to other participants, and only the entities with valid invitation code can join a particular Quorum blockchain system~\cite{Quorum_Blockchain_Service}. In terms of \textbf{\textit{validator selection}}, Quorum supports Proof-of-Authority where the block validators are defined and managed by the deployer. Specifically, Quorum supports alternative consensus protocols, including QBFT, IBFT, Raft, and Clique~\cite{Go_Quorum}. It is noted that \textit{\textbf{block finality decider}} may be different according to the employed consensus protocol. In particular, QBFT, IBFT, and Raft can achieve immediate finality when a new block is appended, whilst forks might occur when Clique is selected.

In addition to the deployer, Consensys (i.e., the project team of Quorum) is also regarded as the \textbf{\textit{benevolent dictator}} in the circumstance that when the deployer or appointed administrator leaves the Quorum system without claiming a new administrator, Consensys can provide support based on the deployment agreement~\cite{Quorum_Blockchain_Service}. Quorum does not implement inherent token issuance or \textbf{\textit{incentive distributor}}, since permissioned blockchain system stakeholders need to adhere to off-chain organisational hierarchy or business agreements where the incentives and decision rights are ascertained. However, the deployer can allocate Ether tokens in a Quorum system if needed~\cite{Go_Quorum}.

In a Quorum system, all activities are recorded and can be used for further analysis and audit via \textbf{\textit{log extractor}}~\cite{Go_Quorum}. The \textit{\textbf{accountability tracer}} is enabled by the generated blockchain account and a public and private key pair~\cite{Quorum_Blockchain_Service}. Considering the above-mentioned \textbf{\textit{participation permission}}, accountability in Quorum can be extended to stakeholders' real-world identity. \textbf{\textit{Protocol upgrade}} in a Quorum blockchain system requires the approval of all participants, then updating the configuration file and restarting the nodes~\cite{Go_Quorum, Besu_upgrade}. Quorum realises \textbf{\textit{network freezer}} by managing the node and account permissioning~\cite{EEA_spe}, suspending the operation of all accounts can freeze a Quorum system. Further, Quorum can specify the types of transactions that a blockchain account is permitted to send~\cite{EEA_spe}. \textbf{\textit{Transaction filter}} is deployed to examine the transaction type.


\subsection{Discussion}

From the evaluation results, it can be found that our reference architecture is correct and usable, as two blockchain system architectures can be mapped on the proposed reference architecture. Comparing Polkadot and Quorum, it is observed that permissionless blockchain systems may implement additional components for voting. The decision rights are allocated to all stakeholders to engage their participation, and increase their trust to the system upgrades, since the results are finalised according to their own choices. Relatively, permissioned blockchain systems like Quorum would centralise the decision rights to the system deployer/administrator, and highlight the permissioning of stakeholders to replicate the real-world positions.

We noticed that several patterns are applied in the off-chain environment, instead of blockchain architectural components. For instance, Polkadot provides a \textbf{\textit{scam list}} introducing several common types of scams to raise stakeholders' awareness~\cite{Polkadot_scam}. Quorum posts an official press release claiming that Consensys acquired Quorum from J.P. Morgan~\cite{Quorum_release}, which can be regarded as an off-chain \textbf{\textit{social contract}}. Besides, since Quorum deploys Ethereum blockchain instances, \textbf{\textit{contract freezer}} is supported by the internal Ethereum virtual machine by default~\cite{pattern_collection}. In addition, we observe more novel governance mechanisms in the evaluation process. For example, Polkadot proposes a concept of \textit{``adaptive turnout biasing"}, where the threshold in a voting process is related to the turnout rate~\cite{Polkadot_design}. Finally, we remark that the proposed reference architecture is adaptive, so that future research can explore more governance patterns and integrate them into this reference architecture.

\section{Conclusion}
\label{sec:conclusion}

Governance is a significant factor throughout the lifecycle of a blockchain system, to ensure normal operation and continuous evolution. Nevertheless, it is found that most existing studies only provide high-level guidelines, while there is a lack of consideration from the respective of architecture design. In this article, we presented a reference architecture to help architects operationalise governance approaches in the future design and development of governance-driven blockchain systems. Specifically, we adopt a widely-accepted blockchain reference model, and apply a set of architectural patterns for governance to the components. We explain the responsibility of each component, and evaluate the correctness and utility of our proposed reference architecture via mapping two existing blockchain system architectures. In future work, we plan to develop decision models for the governance-driven blockchain system design.




\end{document}